C:\SLC4-2P\SLC4v11-2P.doc / 080403

# "Secular Light Curve of 2P/Encke,

# a comet active at aphelion".

Ignacio Ferrín,
Center for Fundamental Physics,
University of the Andes,
Mérida 5101, VENEZUELA
ferrin@ula.ve

Number of pages: 58

Number of Figures: 13

Number of Tables: 1

Proposed Running Head:

**"Secular Light Curve of Comet 2P/Encke"**

Name and address for editorial correspondence:

Dr. Ignacio Ferrín,
Apartado 700,
Mérida 5101-A,
Venezuela,
South America.

email address:

ferrin@ula.ve

## Abstract

We present the secular light curve of comet 2P/Encke in two phase spaces, the log plot, and the time plot.    The main conclusions of this work are: a) The comet shows activity at perihelion and aphelion, caused by two different active areas: Source 1, close to the South pole, active at perihelion, and Source 2, at the North pole, centered at aphelion. b) More than 18 physical parameters are measured from the secular light curves, many of them new, and are listed in the individual plots of the comet.    Specifically we find for Source 1 the location of the turn on and turn off points of activity, $R_{ON}$= -1.63±0.03 AU, $R_{OFF}$= +1.49±0.20 AU, $T_{ON}$= -87±5 d, $T_{OFF}$= +94±15 d, the time lag, LAG(q)= 6±1 d, the total active time, $T_{ACTIVITY}$= 181±16 d, and the amplitude of the secular light curve, $A_{SEC}(1,1)$ = 4.8±0.1 mag. c) From this information the photometric age and the time-age defined in Ferrín (Icarus 178, 493-516, 2005a, and Icarus, 185, 523-543, 2006), can be calculated, and we find P-AGE=97±8 comet years and T-AGE= 103±9 comet years (cy). Thus comet 2P/Encke is an old comet entering the methuselah stage (100 cy < age). d) The activity at aphelion (Source 2), extends for $T_{ACTIVITY}$ = 815±30 d and the amplitude of the secular light curve is $A_{SEC}(1,Q)$ = 3.0±0.2 mag. e) From a new phase diagram an absolute magnitude and phase coefficient for the nucleus are determined, and we find $R_{NUC}(1,1,0)$= 15.05±0.14, and $\beta$= 0.066±0.003.   From this data we find a nucleus effective diameter $D_{EFFE}$ = 5.12(+2.5; -1.7) km.  These values are not much different from previous determinations but exhibit smaller errors.  f) The activity of source 1 is due to $H_2O$ sublimation because it shows curvature.  The activity of source 2 might also be due to $H_2O$ due to the circumstantial situation that the poles point to the Sun at perihelion and aphelion.  g) We found a photometric anomaly at aphelion, with minimum brightness between +393 and +413 days after perihelion that may be an indication of topography.  h) We have re-reduced the 1858 secular light curve of Kamel (1991). There are secular changes in 7 physical parameters, and we achieve for the first time, an absolute age calibration.  We find that the comet entered the inner solar system and began sublimating in 1645±40 AD.   i) It is concluded that the secular light curve can place constraints on the pole orientation of the nucleus of some comets, and we measure the ecliptic longitude of the south pole of 2P/Encke equal to 213.2±4.5º, in excellent agreement with other determinations of this parameter, but with smaller error.   j) Using the observed absolute magnitude of 1858 and 2003 and a suitable theoretical model, the extinction date of the comet is determined.  We obtain ED = 2056±3 AD, implying that the comet's lifetime is 125±12 revolutions about the sun after entering the inner solar system.

## 1. Introduction

There has been little research into the secular light curves (SLCs) of comets, in spite of the fact that they can provide a large amount of information. Besides, the SLCs can be used to create a long term movie of the different apparitions thus compiling the photometric history of the object. In this work we attempt to create two frames of that movie for comet 2P/Encke. The amount of new information generated is very significant, more than 20 new physical parameters.

In a series of papers (Ferrín 2005a, 2006, 2007, from now on Papers I, II, III), we have presented the secular light curves (SLCs) of 9 comets. Recently we have completed work on the *"Atlas of Secular Light Curves of Comets"* (Ferrín, 2008) were 27 cometary SLCs are presented. This work should be consulted to place into perspective the present investigation, and it may be downloaded from the web site cited at the end of this paper. Here we are going to study the SLC of 2P/Encke in detail. This is the comet with the largest number of recorded apparitions, due to its short orbital period of 3.3 years. We also present our own observation of this object in Figure 1.

In Paper I we introduced two views of the SLC, the time plot and the reflected double log plot. Over 18 physical parameters derived from the SLCs are listed in the plots (Figures 2 to 5), many of them new, and provide a wealth of new photometric information on this object that serves to complement other studies. A key to the SLC parameters can be found in Paper I, and an updated key is kept at the web site cited at the end of this paper. The magnitude at $\Delta$, R, $\alpha$, is denoted by V($\Delta$, R, $\alpha$), where $\Delta$ is the comet-Earth distance, R is the Sun-comet distance and $\alpha$ is the phase angle. Briefly the measured parameters listed in the log plot are: the turn on point $R_{ON}$, the turn off point $R_{OFF}$, the asymmetry parameter $R_{OFF}/R_{ON}$, the magnitude at turn on $V_{ON}(1,R,0)$, the magnitude at turn off $V_{OFF}(1,R,0)$, the absolute magnitude before perihelion $m_{1B}(1,1)$, the absolute magnitude after perihelion $m_{1A}(1,1)$, the mean value of both $<m_1(1,1)>$, the slope of the SLC just before perihelion $n_{qB}$, the slope just after perihelion $n_{qA}$, the absolute nuclear magnitude $V_N(1,1,0)$, the phase coefficient $\beta$, the amplitude of the secular light curve $A_{SEC}(1,1) = V_N(1,1,0) - m_1(1,1)$, the effective diameter $D_{EFFE}$, the photometric age P-AGE in comet years (cy). Additionally from the literature the following parameters have been listed in the plot, the perihelion distance q, the aphelion distance Q, the geometric albedo $p_V$, the mean color index $<V-R>$. In the top line of the plot, right side, JF86 means that this is a Jupiter family comet of 86 comet years of age. V.08 is the version of the plot (year), and Epoch 2003 indicates the epoch of the observations that better define the envelope.

The time plot contains a listing of the perihelia plotted, the orbital period of the comet $P_{ORB}$, the next perihelion in format YYYYMMDD, the number of observations used in the plot, the time lag at perihelion, LAG, the turn on time, $T_{ON}$, the turn off time, $T_{OFF}$, the asymmetry parameter, $T_{OFF}/T_{ON}$, the active time, $T_{ACTIVE}$, the slope at turn on, $S_{ON}$, the slope at turn off, $S_{OFF}$, and the time-age, T-AGE, in comet years (cy).

*The importance of these secular light curves is that they show not only what we know, but also what we do not know, thus pointing the way to meaningful observations.* The SLCs are very useful for planning observations. For example from the time plot we

can deduce that the bare nucleus of 2P/Encke can be seen only at times with respect to perihelion, $-228 < \Delta t < -92$ d, $+94 < \Delta t < +161$ d, and $+393 < \Delta t < 413$ d (confirmation Figure 4 and 5). Beyond these intervals the nucleus is coma contaminated.

Paper I justifies the procedures adopted in this paper, and reading it would help to get acquainted with the plots, procedures and parameters measured from them. It also helps to place into perspective the results found in the present work.

The photometric system used in this work is described in Paper I but a short description follows. No corrections were applied to the observations. The methodology outlined in that paper makes no corrections and lets the brightest observations define the envelope. It was found and justified there, that the envelope of the observations is the best representation of the SLC of comets. The justification is that all visual observations of comets are affected by several effects, all of which decrease the perceived brightness of the object by washing out the outer coma: moonlight, twilight, haze, cirrus clouds, dirty optics, excess magnification, large aperture, etc. *There are no corresponding physical effects that could increase the perceived brightness of a comet.* Thus the envelope is the correct interpretation of the SLC and it does not require corrections of any kind. This rule is confirmed observationally. The top observations in Figure 2 and 3 show a sharp edge, while the bottom observations have no edge and are distributed in a diffuse area.

After presenting the evidence, we will conclude that 2P/Encke is an old comet entering the methuselah stage (100 cy $<$ P-AGE), active at perihelion (Source 1) and with remarkable activity centered at aphelion (Source 2). In the context of this investigation *"a source"* may include several active regions lumped together rather than a surface sublimating on every square inch. Since aphelion takes place at Q= 4.09 AU the question is what drives the sublimation. We will conclude that $H_2O$ may be the source of this activity due to a circumstantial situation: a pole of the comet points to the Sun at perihelion and aphelion. Additionally it has been possible to achieve an *absolute age calibration and to determine the extinction date*.

## 2. Observations

Our observational program of faint comets was established in the 90s, and comet 2P/Encke was imaged on November 18[th], 1993, using the 1 m f/3 Schmidt telescope of the National Observatory of Venezuela. The exposure time was 420s and the night was photometric. The image is shown in Figure 1 and the field size is 5'x6.7'. The telescope was coupled with a Thompson 7883 CCD, of 576x384 pixels. Since this detector had a fairly low quantum efficiency, the observations were filterless in order to maximize the S/N ratio. The maximum response curve between 6500 and 7500 A, is approximated fairly well by a broad band R filter. The CCD is quite blind at 4500 A. Processing was done using standard reduction procedures. Calibration of the magnitudes was done using the R magnitudes of the USNO A2.0 catalog. To avoid errors from poor magnitude stars, a least squares calibration of the field was done with no less than 15 stars. Stars deviating from the linear fit by more than $3\sigma$ were dropped. This calibration method is good to ~±0.1 magnitudes. The comet was then at an Earth's distance $\Delta= 0.94$ AU, sun distance R= 1.60 AU, and phase angle, $\alpha = 35.6°$, $\Delta t = -83$ d.

We found the following results, $R(\Delta,R,\alpha)$= 17.0±0.15, $R(1,1,\alpha)$= 16.11±0.15, $V(1,R,0)$= 15.08±0.15.   In Paper II we have compiled all the nuclear color observations of this comet, and we find a mean color index < V - R > = 0.44±0.04, where <x> denotes the mean value.    We can then calculate that $V(\Delta,R,\alpha)$= 17.44±0.15.   These values are plotted in the SLC (Figures 2 and 3) and by good luck lie in the region were the comet is turning on, thus helping to define the extent of activity of 2P.    In fact the observation was made only +4 days after turn on of activity.

## 3. 2P/Encke

Lamy et al. (2005) have a very useful compilation on physical properties of comets, and it is recommended reading on other physical parameters of this comet.

*Data Sets*    We were able to extract 768 observations of this comet from the literature. The literature is very extensive because of its short orbital period of 3.3 years and because of its favorable observing conditions.    We will restrict ourselves to those works dealing with photometry and related matters (pole orientation). Up to now all the photometric observations available had not been placed into one single plot. Observations by the following authors were used: Ney (1974), Jewitt and Meech (1987), Luu and Jewitt (1990), Fernandez et al. (2000), Meech et al. (2001) (published in Belton et al. (2005)), Lowry et al. (2003), Belton et al. (2005), Fernandez et al. (2005), Hergenrother et al. (2006), Lowry and Weissman (2007).  For visual observations we used the dataset compiled in the International Comet Quarterly (Green, 2006).  Also, there are many visual and CCD observations available in the web.  One remarkable data set is that of Biesecker (2000) comprising total magnitudes derived from SOHO spacecraft images taken at perihelion.

The dataset of Fernandez et al. (2005) is remarkable in its extent and precision. It contains the data presented separately by Lowry et al. (2003) and Hergenrother et al (2006).   Because it has been taken when the comet was at aphelion, and because the spacing between sets is as small as a day, this data set deserves additional analyzing to determine the day to day variability, and this is done in a dedicated section below.

*Secular Light Curves*    Using the reducing procedure described in paper I, the SLCs were plotted in Figures 2 to 5. As explained in the introduction the SLCs are presented in two views, the Log plot, and the time plot.   The importance of the log plot is that power laws on R ($R^n$), plot as straight lines.   The log plot is a double logarithmic plot reflected at R=1 AU to allow the determination of the absolute magnitude by extrapolation to Log R= 0. Time runs on the horizontal axis toward the right, although non-linearly.   On the other hand, the importance of the time plot is that time runs linearly in the horizontal axis, showing the brightness history of the object.   The two plots complement each other, and both are needed because they provide two independent views of the same phenomena on different phase spaces and give different physical parameters.   Comet 2P/Encke requires an additional plot to show the activity at aphelion (Figure 4).

The top observations in Figures 2 and 3 are mostly visual measurements by amateurs, denoted by $m_1$, while faint observations are nowadays CCD measurements by amateurs and professionals, denoted by V.    Ferrín (2005b) has found a negligible correction from one system to the other, $V - m_1$ = -0.026±0.007 magnitudes.  This result can be checked. For stars Zissell (2003) finds V = $m_1$ - 0.182 * (B - V ) + 0.07. Since comets have a mean V - B = 0.5, then V = $m_1$ - 0.02, in perfect agreement with our result.

The log plot (Figure 2) represents a normal SLC, with a well defined turn on and turn off points, and a slight time lag at perihelion. The time plot (Figure 3) also shows normal activity near perihelion (Source 1). Both plots allow the determination of over 18 physical parameters, many of them new, and they are listed in the plots.

From Figure 2, the log plot, we can determine that the comet turns on at $R_{ON}$ (S1)= -1.63±0.03 AU and turns off at $R_{OFF}$ (S1) = +1.49±0.2 AU (the minus sign indicates observations prior to perihelion). We also measure the amplitude of the secular light curve, $A_{SEC}(1,1)$= 4.8±0.1, and the mean absolute magnitude of the comet <$m_1(1,1)$> = 10.7±0.2. However it must be pointed out that the SLC is asymmetric and thus the absolute magnitude before perihelion ($m_{1B}(1,1)$= 9.8±0.1) is quite different from the absolute magnitude after perihelion ($m_{1A}(1,1)$= 11.5±0.1), presenting a decay of 1.7±0.15 magnitudes at R= 1 AU from the Sun post perihelion.

The time plot, Figure 3, also gives some interesting information. We find that the comet turns on at $T_{ON}$= -87±5 d, and turns off at $T_{OFF}$= +94±15 d, while the time lag at perihelion is LAG(q)= 6±1 d. We have not attempted to reconcile the values derived from the two plots, to give some idea of the errors involved and to provide two independent determinations of the onset and offset of activity. Another parameter is the total active time of the comet $T_{ACTIVITY}$ = 181±15 d or ~6 months. Additional parameters are listed in the plots.

One remarkable characteristic of this comet, it that it exhibits activity at aphelion without a detectable coma (confirmation Figures 2, 4 and 5). The aphelion activity was recognized by a number of researchers, Sekanina (1976), Barker et al. (1981), Meech et al. (2001), Meech et al. (2004), Belton (2005). The SLC shows that the observations are too bright above the nuclear line ($R^{*2}$). In order to visualize the activity more clearly, we have compiled and plotted all observations near aphelion, and they are shown in Figure 4. It is remarkable that the activity is quite significant in time interval ($T_{ACTIVITY}$ = 815±30 d) and in intensity ($A_{SEC}(1,Q)$ = 3.0±0.2 mag). For comparison, the total orbital time of the comet in 2003 was only 1206 days. Below we will suggest an origin for the activity for Source 2.

Additional interesting parameters deduced from Figure 4 are the turn on and turn off distances of Source 2. We find $T_{ON}(S2)$= +160±25 d, which corresponds to a solar distance $R_{ON}(S2)$= +2.41 AU, and $T_{OFF}(S2)$= +975±15 d, which corresponds to $R_{OFF}(S2)$= -2.94 AU. Additional parameters of S2 are the time lags measured from perihelion LAG(q) = +718±30 d and from aphelion LAG(Q) = +120±30 d.

*Photometric anomaly*    Figure 4 shows a minimum brightness at ~+400 d. In Figure 5 the minimum of activity at aphelion is enlarged. We see a linear decay in brightness between +312 and +393 d after perihelion, and a minimum between +393 and +413 d. In future apparitions it would be interesting to verify if the minimum repeats in the same place. This might be an indication of *topography or the turn off of and active region.* To be able to follow a specific point on the surface of a comet is of great interest, because it may allow a more precise determination of rotational evolution and of excited rotational periods.

*Photometric age P-AGE and time-age T-AGE* In papers I and II the following age-related parameters were defined: the photometric age and the time-age,

$$P\text{-}AGE = 1440 / (R_{ON} + R_{OFF}) * A_{SEC}$$

$$T\text{-}AGE = 90240 / (T_{ON} + T_{OFF}) * A_{SEC}$$

and we have concluded that they are very robust definitions. Comet ages have been scaled to human ages. Using the information derived above, we find P-AGE = 97±8 cy (comet years) and T-AGE= 103±9 cy. If we define 75 cy < old < 100 cy, and 100 cy < methuselah, then this comet is going from old to methuselah. It must be emphasized that what P-AGE and T-AGE really measure is *activity as a proxy for age* and that is the reason they are measured in comet years.

*Geometric albedo* An important physical parameter needed to derive the dimension of the nucleus is the geometric albedo in the R-band, $p_R$. The only determination of this parameter is by Fernandez et al. (2000) who find $p_R$ = 0.047±0.023. No doubt this is a difficult measurement, so the error is not surprising. Thus we would benefit from a re-measurement, and this is best done when the bare nucleus is visible ( -228 d < $\Delta t$ < -92 d, +94 d < $\Delta t$ < 161 d, and +393 d < $\Delta t$ < 413 d , confirmation Figure 4).

*Phase curve, phase coefficient, and absolute nucleus dimensions* The best way to derive the absolute nuclear magnitude of a comet is by plotting the phase diagram, V(1,1,$\alpha$) vs $\alpha$. The slope of this curve gives the phase coefficient, $\beta$, which is needed to reduce nuclear observations to a common phase angle of 0º. Fernandez et al. (2000) have done this type of plot. However these authors included photographic observations, and coma contaminated values are clearly seen in their phase curve. The CN line at 3883 A is the most intense of the whole cometary spectrum, thus many of those measurements are contaminated by the intense CN blue coma. From the plot they find $R_N(1,1,0)$= 15.2±0.5 and $\beta$= 0.061 mag/º (no error quoted). So it is of scientific interest to use only modern CCD observations (sensitive to the red part of the spectrum) to repeat the phase plot and find if there is any difference.

The new plot shown in Figure 6 contains only observations carried out with CCDs, IDS (image dissector scanner) and HST (Hubble Space Telescope) that have already been calibrated and instrument corrected (Fernandez et al., 2000). We find the following results: a) The absolute nuclear magnitude is $R_N(1,1,0)$= 15.05±0.14 and $\beta$= 0.066±0.003 mag/º. The new absolute magnitude is in close agreement with the one derived by Fernandez et al. (2000), but the errors have been reduced significantly.

Why should these values be better than previous determinations? Because we have imposed to the phase plot three additional constraints: 1) That the observations be devoid of coma (this means to drop photographic observations). 2) That the observations lie inside the rotational amplitude of the nucleus. And 3) that the absolute magnitude $V_N(1,1,0)$ be in agreement with the faintest nuclear observations at aphelion (confirmation Figure 5). As a result, the errors are smaller.

In Paper II we found a mean color index < V - R > = 0.44±0.04, and with this information the absolute magnitude in the visual is $V_N(1,1,0)$= 15.49±0.15.   When this value is plotted in Figures 4 and 5, the nuclear line is in accord with the lowest observations made by Fernandez et al. (2005) and Meech et al. (2001) at aphelion thus satisfying constraint 3) listed above.   As a byproduct we find the *peak-to-valley* amplitude of the *rotational* light curve at aphelion, $A_{PTV}$ = 0.62±0.08 magnitudes.

In Figure 6 we have also plotted our observation of 2P/Encke.   In spite of having been taken only +4 days after turn on, it is already coma contaminated.   This result shows that the phase diagram is a very sensitive way to detect coma activity and justifies the adoption of constraint 1) cited above.  Constraint 2) justifies itself.

A compilation of phase coefficients, $\beta$, has been presented in Paper I, and it can be seen that this value for 2P/Encke is one of the largest, implying an extreme surface texture (a conclusion previously reached by Fernandez et al. (2000) and Lamy et al., (2005)).

*Nucleus dimensions*     This comet has been observed by radar (Kamoun et al., 1981), and these researchers have obtained a nucleus diameter of D= 3.0 (+4.6, -2.0) km. Fernandez et al. (2000) obtain D= 4.8±0.6 km. Reach et al. (2000) find D= 4±2 km. Lowry et al. (2003) find D= 8.1±0.12 km.  Kelley et al. (2006) find two values, D= 4.68±0.28 km, and D= 3.44±0.20 km, but consider the possibility that the second one may be evidence of a non-spherical nucleus, and deduce a ratio of axis a/c > 1.4 .   Harmon and Nolan (2005) had a second chance to see the comet by radar during the close approach of 2003, and they find 4.84 km < D < 7.44 km.

We can determine the effective diameter $D_{EFFE}$, using the absolute nuclear magnitude $R_N(1,1,0)$ from the phase plot (Figure 4), the geometric albedo in the red determined by Fernandez et al. (2000), $p_R$, and the amicable formula deduced in Paper III:

$$\text{Log} [ p_R D^2_{EFFE} / 4 ] = 5.510 - 0.4 R_N(1,1,0) \qquad (1)$$

where $R_N(1,1,0)$ is the absolute nuclear magnitude in the R band. We find $D_{EFFE}$ = 5.12(+2.5;-1.7) km.  The error is mostly due to the error in the geometric albedo which is rather large. This value is consistent with previous determinations but should be nearer to the truth because of the imposed constraints 1) to 3).

As an example of why the determination of the nuclear size benefits from the information given by the SLCs, let us consider the recently derived size by Lowry and Weissman (2007) who find a mean effective diameter = 7.90±0.12 km which is surprisingly large and accurate.  This result contains three inaccuracies.  The first one arises because these authors use the geometric albedo derived by Fernandez et al. (2000), $p_R$ = 0.047 forgetting to include the associated error of ±0.023 discussed above in the geometric albedo section.  Thus their propagation or errors is incorrect resulting in an unrealistically small error of ±0.12 km.   The second inaccuracy arises because they use a phase coefficient of $\beta$ = 0.06 mag/º.  However this value cited from Fernandez et al. (2000), contained photographic magnitudes as well as total magnitudes above the rotational amplitude. Our result presented above in the phase plot section, shows that the correct phase coefficient is β = 0.066 ± 0.003 mag/º (confirmation our Figure 6). The third inaccuracy appears because they assume that the nucleus is devoid of an unresolved coma. To verify this let us take the photometry of the first night, October 3rd, 2002. They find the mean night value for the photometry $< R(\Delta, R, \alpha) > = 19.78 \pm 0.03$. Since this observation was made +754 days after perihelion at a phase angle of 7.12º, and using a color coefficient from these authors of $< V - R > = 0.39 \pm 0.06$, we find a reduced magnitude V(1,1,0) = 14. 34 ± 0.08. Plotting this value in Figure 4 that shows the aphelion activity of comet 2P/Encke, we find that the observation was actually 1.15 mag above the nucleus due to an unresolved coma. No wonder then that the radius they found was so large.

Alternatively the unresolved coma could have been detected from the phase plot, and this is confirmed in Figure 6 were we see that it lies well above the rotational amplitude.

The moral of this tale is clear. *Erroneous values can be derived by ignoring the information given by the secular light curve.* This is further proof of the importance and significance of the concept of the SLCs derived in these series of works.

*Rotational period* Another surprising result obtained by these authors is that their preferred rotational period is 11.083±0.003 h. However the associated rotational light curve exhibits only one maximum and one minima, something impossible for a rotating ellipsoid. In fact their alternative period 22.2190 h exhibits two maxima and two minima separated by 180º, which is what is expected of a rotating ellipsoid. Since this analysis contains the lengthy photometry by Fernandez et al. (2005) it is tempting to conclude that this is the real rotational period of the comet. And this is confirmed by the fact that their October 3 and 4 photometry fits perfectly with the rotational light curve if it is displaced upward by only 0.25 magnitudes.

*Aphelion activity without coma* 2P/Encke presents double activity, at perihelion and aphelion (Figures 4 and 5). The list of authors that have contributed to this knowledge has been given above. The activity at aphelion extends for 815±30 days, while that at perihelion lasts for only 202±15 days. This is a most unusual behavior not shown by any other comet up to now, and suggests that two sources are active. Since activity at aphelion takes place at 4.09 AU from the sun, the question arises of what substance is controlling sublimation. According to a calculation by Delsemme (1982) it was previously believed that water ice sublimation could not be sustained beyond 3 AU. However recent calculations by Meech and Svoren (2005) show that water ice sublimation can be sustained up to 4-6 AU.

Sekanina (1988a) and Festou and Barale (2000) find that one pole of the comet points to the sun at perihelion. This opens the possibility that the activity at aphelion is produced at the other pole, in which case 2P/Encke shows activity at the two poles, a most peculiar situation.

**4. Dataset of Fernandez et al. (2005)**

It is not clear if the activity at aphelion is sustained (continuous) or impulsive (superposition of discrete outbursts). The data set of Fernandez et al. (2005) of *rotational* measurements (Table 1), is remarkable for its extent and precision, and because they were carried out at aphelion, they can be used to say something about the type of activity exhibited at those distances.

Means and ranges have been calculated for every night, and plotted as a function of time in Figure 7. There are four observing intervals. In the first, third and fourth intervals the comet remains practically constant. The second interval in the middle lasting 15 d is much fainter than the other three and exhibits a drop of 1.1 magnitudes in 13 days. The lowest night is 1.5 magnitudes below the steady state nights. Notice also that the amplitude of the rotational light curve is larger for the fainter magnitudes, implying that brighter magnitudes are coma contaminated. This is the justification for imposing constraint 1) in the phase plot.

The following is evidence of *sustained activity* (Figure 7): a) From July 19[th] to August 13[th], an interval of 25 days, the comet maintained its mean brightness within ±0.1 magnitudes. b) During September 12[th], 2002, and September 16[th], an interval of 6 days, the comet maintained its mean brightness within ±0.03 magnitudes.

The following is evidence of *impulsive activity*: a) A decrease of 1.1 magnitudes in an interval of 13 days, from September 23[r] to October 6[th], 2001. b) What looks like a linear decrease of 1.3 magnitudes in an interval of 82 days, seen in Figure 5 from $\Delta t$ = +310 to +392 d after perihelion, producing a sharp minimum shown expanded in Figure 5.

From Figures 4 and 5 we conclude that the comet exhibits both types of behavior. The impulsive activity is reminiscent of the one exhibited by comets 29P/Schwassmann-Whachmann 1 and 95P/Chiron.

## 5. Pole Orientation and the SLC

The shape of the SLC is determined by several physical parameters that are entangled: the pole orientation (I, $\Phi$), the composition, and the thermal conductivity. These are the first order parameters. There may be second order parameters (like topography and shape of the nucleus) that affect the SLC, but they will not be considered because they produce a much smaller influence. The question is how to disentangle them. There are two cases in which it may be relatively simple: pole perpendicular to the orbit, and pole parallel to the orbit.

Consider the case where the pole is perpendicular to the orbit. The Sun shines all the time on the equator of the comet. Thus the parameter LAG (measured at the maximum brightness of the SLC) must be entirely due to the thermal lag. However Lamy et al. (2003) have made a first order calculation assuming a thermal inertia of 10 $J/K/m^2/s^{1/2}$, and found that the heat wave penetrates rather slowly, approximately 1 cm/day. This value is very small, and it implies that the observed LAG time in the SLC could not be due to a thermal lag.

Consider now the case where the pole is parallel to the orbit. Then the parameter LAG must take place when the pole is pointing to the Sun, and minimum light may take place when it is shining on the equator.  This may be the case of 2P/Encke, with the peculiarity that one pole points nearly to the sun at perihelion.

The situation for 2P/Encke is shown in Figure 8, where the points of turn on and turn off of the two active sources (S1 and S2) have been indicated and are taken from Figures 2, 3 and 4.   This Figure is actually saying something about the distribution of volatiles on the surface of 2P/Encke, and this distribution is shown in Figure 12.

*Pole orientation*    Sekanina (1988b) and Festou and Barale (2000) have obtained a solution for the spin axis that is practically the same (see also Samarasinha, 1997).  The pole points to the Sun 4 days after perihelion according to the Sekanina's solution, and 6 days for the Festou and Barale's solution.    These values are in excellent agreement with the LAG time deduced from the SLC, 6±1 days (confirmation Figure 2) because maximum brightness corresponds to maximum temperature.

It is thus possible to derive the pole orientation solely from the SLC.  If we assume that the maximum brightness of the SLC corresponds to the moment in which the comet points one pole to the sun, then a LAG= +6±1 d can be translated into an angle $\nu$ = 26.7±4.5º for the true anomaly.    The SLC can constraint the *ecliptic longitude* of the pole.

Sekanina (1988b) concludes that comet 2P/Encke has a pole pointing to ecliptic longitude 200º, ecliptic latitude +18º.    From Figure 8 and from the fact that the longitude of perihelion of this comet is $\varpi$ = 186.5º , we find that the pole of this comet points toward ecliptic longitude 213.2±4.5º.    Thus our observational value is in close agreement with Sekanina and the errors superimpose.   The result is also in agreement with Festou and Barale (2000).     The SLCs can also derive a value for the *ecliptic latitude* through modeling of the light curve, but this is beyond the scope of this paper.

This implies that the SLCs derived in Papers I to III can be used to place limits on the pole orientation of those comets.   For example, the extreme LAG= +155±10 d derived for comet 133P/Elst-Pizarro (Paper II) may indicate that the pole of this comet points to the sun at that time.

The south pole of 2P/Encke almost points to the Sun at perihelion.   It is interesting to notice that this result supports the conclusion reached by Samarasinha (1997) who found that pole orientations tend to migrate spiraling until one pole points to the Sun at perihelion.  In the case of 2P/Encke, it is the south pole.

## 6. Secular evolution, 1858 vs 2003.

Kamel (1991) has determined the secular light curve of 2P/Encke in 1832-1871 (effective epoch 1858).    Although he concludes that there is no significant secular change, his plots and our plots say otherwise (confirmation Figure 9). We re-reduced his data (Kamel, 1992) and the new log plot for 1858 appears in Figure 9 and the time plot in Figure 10.    Using our interpretation of the envelope as the correct description of the SLC (paper I), we find that there are at least 7 physical parameters that have shown temporal evolution, LAG, $m_{1B}(1,1)$, $m_{1A}(1,1)$, $R_{ON}$, $R_{OFF}$, P-AGE and T-AGE:

a) There is a shift in LAG. The maximum brightness in 1858 took place with LAG= -14±2 d (before perihelion), and LAG= +6±1 d in 2003, an interval of 145 years. So peak brightness shifted by +0.14 d/y or 0.46 d/apparition in this interval.

b) Change in absolute magnitude m(1,1). In 1858 R= 1 AU takes place at -43.3 d. Using Kamel's Figures 9 and 10, we find the comet had an absolute magnitude before perihelion $m_{1B}(1,1)$= 7.5±0.2 in 1858, while in 2003 we find $m_{1B}(1,1)$= 10.0±0.2. After perihelion we find in 1858 $m_{1A}(1,1)$= 10.7±0.2, and in 2003 $m_{1A}(1,1)$= 11.5±0.2. The comet is getting fainter in absolute magnitude. The asymmetry at R= 1 AU was also larger, 3.2 magnitudes in 1858 vs 1.5 mag in 2003, *half the value*. The decaying rate of $m_{1B}(1,1)$ has been 0.045 mag/apparition, while that of $m_{1A}(1,1)$ is 0.018 mag/apparition. A decay in absolute magnitude with time is also exhibited by the SLCs previously published.

c) The maximum peak brightness has changed from $m_{MAX}(1,LAG)$= 5.8±0.2 in 1858 to $m_{MAX}(1,LAG)$= 6.2±0.2 in 2003, a statistically significant change.

d) In 1858 the turn on point was $R_{ON}(1858)$= -2.18±0.10 AU, while in 2003 we find $R_{ON}(2003)$= -1.63±0.03. There is a significant decrease in turn on distance in the interval of 145 years, at the rate of -0.0127 AU/apparition.

e) In 1858 the turn off distance was $R_{OFF}(1858)$= +1.73±0.10 AU, while in 2003 we find $R_{OFF}(2003)$= +1.49±0.2 AU, implying a decrease of -0.0057 AU/apparition. Thus the SLC has been narrowing with time, a behavior also shown by the comets presented in papers I to III.

f) However the most interesting change is in P-AGE and T-AGE. We find P-AGE(1858)= 58±3 cy and T-AGE(1858)= 61±4 cy. These values correspond to a middle age comet. Since we also know these ages in 2003, it is possible to calibrate P-AGE [cy] and T-AGE [cy] *in absolute Earth years* [y] to obtain the current date. This calibration is shown in Figure 11.

CD  = 1642 + 3.72 * P-AGE

CD  = 1647 + 3.45 * T-AGE

where 1642 AD and 1647 AD are not the birth year of the comet but *the year in which the comet began sublimating*. Thus we deduce a zero age date ZAD = 1645 ± 40 AD (Figure 11). However since sublimation is related to solar distance, *this must also be the year in which the comet entered the inner solar system.* The conversion factors 3.72 and 3.45 come in units of y/cy and give the change of scale from comet time to earth time.

*Lifetime*. We are interested in knowing the total lifetime of the comet in number of revolutions around the sun. Several estimates have previously been made. If we scale Weissman (1980) results to Encke's values, we find a timescale of the order of a few hundreds orbits. Rickman et al. (1991) find that small nuclei can be consumed on a short timescale (<100 revolutions). Cowan and A'Hearn (1979) published a sublimation model and calculated the lifetime in number of orbits for 2P/Encke. For a Bond albedo of 0.4, they find a timescale of ~250 orbits. However their albedo is previous to the comet Halley encounter, and now much smaller values are favored.  So the new estimate is of the order of ~150 orbits.

To make our own estimate of the lifetime, we will make use of the secular absolute magnitudes reported by Ferrín and Gil (1988).  These authors have reduced historical absolute magnitudes using visual observations and observations made with small instruments.  However their dataset shown in Figure 13 has to be moved by -0.6 magnitudes to take into account that they took the mean value of the observations, while we now know that the correct interpretation is the envelope.  Additionally we will adopt the theoretical model there developed for a comet that looses a constant layer of water ice per apparition.

Data and three models are compared in Figure 13.  We see that the sublimating model fits the observations rather well.  A model based on extinction due to suffocation would decrease the absolute magnitude asymptotically and thus would not have a sharp end.  This alternative model can not be ruled out, but the sublimating model is useful because it predicts an accelerated decay, and this may be evident in just a few decades, since the predicted extinction date is just ED= 2056±3 AD.   This implies a total number of revolutions of 125±12 in agreement with previous estimates.

For other comets the timescale may be much larger, but for 2P/Encke the observational value may be correct because of three reasons: a) The perihelion distance of this comet is very small, q= 0.33 AU, and compared with a comet at q=1 AU, 2P receives 9 times more solar energy and thus ages 9 times faster.    b) The pole points to the sun at perihelion and aphelion.   Thus at perihelion the comet is in daylight all the time.  The surface temperature is much higher than for a rotating comet with a pole perpendicular to the orbit, in which case the solar energy is spread out over the whole equatorial region.    Thus sublimation is optimized.   Finally, c) the comet is active at aphelion where other comets are inactive.  a) to c) conspire to produce aging at a larger rate than for other Jupiter family comets and thus produce a shorter lifetime.

*Orbit and origin*   The orbit of this comet is very odd, being completely decoupled from Jupiter.  Levison et al. (2006) have tried to explain this condition by integrating a large number of test particles in similar orbits.  They found that it takes roughly 200 times longer to evolve onto an orbit like this than the typical cometary physical lifetime.   To reconcile this result, they propose that:  (a) 2P/Encke became dormant soon after it was kicked inwardly by Jupiter; (b) it spent a significant amount of time inactive while rattling around the inner solar system; or (c) it only became active again as the $\nu_6$ secular resonance drove down its perihelion distance.

We will define a *normal comet* as one that exhibits sustained activity quasi-symmetric with respect to perihelion, with a well defined turn on point before perihelion, and a well defined turn off point after perihelion (confirmation paper I).  After comparing with the other comets presented in papers I to III, it can be concluded that this comet is entirely normal, not showing any characteristic that could suggest a period of inactivity.  On the contrary, it shows activity at aphelion at 4.09 AU from the Sun.  Thus we have not been able to find any evidence whatsoever in the SLCs, to support conclusions (a) to (c) above.  In fact, from what we have been able to learn from papers I-III, those conclusion are untenable because there is physically no way that a comet may rattle around the inner solar system without exhibiting a strong cometary activity.  That situation has never been observed for any comet.  The $\nu_6$ resonance is placed at ~2.05 AU from the Sun (Scholl and Froeschle, 1991) and thus is located well inside the region of activity shown by the SLC in 1858 AD, and could not be a hiding place for activity.   Our secular light curve constraints the residence orbit to beyond ~5-6 AU where it would show no activity and thus where it could remain for a considerable time without aging.

Another dynamic calculation has been carried out by Pittich et al. (2004).  They get a much smaller timescale including in the calculation non-gravitational forces.  However, according to Levison et al. (2006) (who did not include non-gravitational forces in their calculation), the non-gravitational forces used in their model are unrealistically large and were held constant.   Since the SLCs show evidence of a short lifetime the question of the origin of the orbit of comet 2P/Encke remains without satisfactory solution and is still a matter of debate.

*We are fully aware that currently the zero age date, ZAD = 1645₆40 AD,  can not be explained dynamically.  But neither can the current orbit.*  Additionally the age of the meteor stream associated with this comet is ~4700 y or ~1400 orbits about the sun.  Thus one is tempted to conclude that the comet was inactive for a long time.  But this situation can not be explained either.   Thus the mystery deepens. *Such is the substance of science.*

To conclude this issue, we must mention that Whipple and Hamid (1972) carried out an intensive search for Encke's comet among nearly 600 transient objects in Ho's (1962) catalog, which covers observations until the period AD 1600.  They turned up a handful of suspects but no definitive candidate.  This result is in perfect agreement with our determination, specially if we take into consideration that before 1600 AD, the absolute magnitude of the comet should have been brighter than 5.5 (confirmation Figure 13), and thus it is inconceivable that it should have escaped detection at times passing near the Earth.

## 7. Conclusions

The main conclusions of this work are:
1) We define a *normal comet* as one that exhibits sustained activity quasi-symmetric with respect to perihelion, with a well defined turn on point before perihelion, and a well defined turn off point after perihelion (confirmation paper I).   Using this definition we find that 2P/Encke is a normal comet.

2) From the secular light curves we measure the following parameters for the perihelion activity or coma phase (Source 1): The location of the turn on and turn off points of activity are $R_{ON}$= -1.63±0.03 AU, $R_{OFF}$= +1.49±0.20 AU, $T_{ON}$= -87±5 d, $T_{OFF}$= +94±15 d, the time lag, LAG(q)= 6±1 d, the total active time, $T_{ACTIVITY}$= 181±16 d, and the amplitude of the secular light curve, $A_{SEC}(1,1)$ = 4.8±0.2 mag.

3) From this information the photometric age and the time-aged defined in papers I and II, can be calculated.  Comet 2P/Encke is an old comet, with P-AGE=97±8 comet years, and T-AGE= 103±9 comet years.

4) The activity at aphelion (Source 2), extends from $T_{ON}$= +160±25 d to $T_{OFF}$= +975±15 for a total active time $T_{ACTIVITY}$ = 815±30 d and amplitude of the secular light curve $A_{SEC}$ (1,Q) = 2.7±0.2 mag.    We also find the peak-to-valley amplitude of the rotational light curve at aphelion, $A_{PTV}$ = 0.62±0.08 magnitudes.

5) We create a new phase diagram based on a reanalysis of older observations, to which 3 new constrains are imposed: no photographic observations used because they are contaminated by the 3883 A CN line, no observations used beyond the amplitude of the rotational light curve, and agreement with aphelion observations where the nucleus should be more visible.    We find an absolute magnitude and phase coefficient $R_N$(1,1,0)= 15.05±0.14, and β= 0.066±0.003.     From this data we find a nucleus effective diameter $D_{EFFE}$ = 5.12(+2.5;  -1.7) km not different from previous determinations but with smaller error and satisfying three additional constraints (see text).

6) The activity of source 1 is due to $H_2O$ sublimation because it shows curvature.  The activity of source 2 might also be due to $H_2O$ due to the circumstantial situation that the poles point to the Sun at perihelion and aphelion.

7) It is concluded that the secular light curve can place constraints on the pole orientation of the nucleus of some comets, and we measure the ecliptic longitude of the south pole equal to 213.2±4.5º, in excellent agreement with the values from Sekanina (1988b) and Festou and Barale (2000), but with smaller error.

8) We obtained the SLC for 1858 using data from Kamel (1991).  We find that at least 7 physical parameters have changed since 1858.  In particular the photometric age and time-age were found to be P-AGE(1858)= 58±3 cy and T-AGE(1858)= 61±4 cy much younger than current values.   This allows the determination of the date of insertion of the comet into the inner solar system and start of activity as 1557±90 AD.   Although this timescale for evolution is shorter than previously believed, it is justified in the text.

9) It is interesting to follow the comet at aphelion closely to verify if the linear decay in brightness between Δt = +310 and +393 d after perihelion repeats in the same place at other apparitions. This could be an indication of topography.  It could also be interesting to obtain time series differential photometry of the nucleus at the time of minimum (Δt= +393 to 412), because this seems to be the absolute minimum brightness of the whole orbit.

10)  We conclude that 2P/Encke shows secular evolution in seven parameters.  The SLC has been shifting in LAG, absolute magnitude, turn on and turn off distances, age, and it is changing shape.    For the first time we have been able to achieve an absolute calibration of age and it was found that the residence time inside the inner solar system is small, but in accord to several calculations. It is also in accord with the small perihelion distance and peculiar pole orientation of this comet that produce enhanced aging.

11) A reinterpretation of the rotational light curves obtained by Lowry and Weissman (2007) allows to conclude that the best rotational period for the comet that also fulfills the lengthy photometry of Fernandez et al. (2005), is 22.2190 h.

12) Using secular information on the absolute magnitude and a theoretical model based on the sublimation of volatiles, it is possible to predict the extinction time. We find ET= 2056±3 AD, implying that the comet made 125±12 returns after entering the inner solar system. This value is in accord with previous estimates.

13) This comet will have a close approach to Earth in October 17[th], 2013 at the comet-Earth distance $\Delta$ = 0.478 AU, when the object will be within reach of radio telescopes and will be a good radar target (Harmon et al., 1999; Harmon and Nolan, 2005).

The *"Atlas of Secular Light Curves of Comets"* as well as tables of cometary physical properties is available and are regularly updated at the web site: http://www.webdelprofesor.ula.ve/ciencias/ferrin.

## 8. Acknowledgements

To Oliver Hainaut and to an anonymous referee for their careful reading of the manuscript that resulted in significant improvements of its scientific value. To the Council for Scientific, Technologic and Humanistic Development of the University of the Andes for their support through grant number C-1281-04-05-B. The observations described in this work were carried out at the National Observatory of Venezuela (ONV), managed by the Center for Research in Astronomy (CIDA), for the Ministry of Science and Technology (MCyT). To the TAC of CIDA for time granted to observe these faint objects. To the night assistants, Freddy Moreno, Ubaldo Sanchez and Orlando Contreras for their help at the telescope. And to Franco della Prugna, Gerardo Sanchez and Gustavo Sanchez for their technical support at the telescope.

Table 1. Data sets of Fernandez et al. (2005) (1 to 17) and Lowry and
Weissman(2007) (18 and 19). Nightly means and ranges ($A_{PTV}$). $\beta$ = 0.066 mag/°

| # | YYYY | MM | DD | $\Delta t[d]$ | $<m(R)>$ | $m(V)$ | $A_{PTV}$ | $\Delta[AU]$ | $R[AU]$ | $\alpha$ | $V(1,1,\alpha)$ | $V(1,1,0)$ |
|---|------|----|----|------|---------|--------|-------|---------|--------|-------|-----------|-----------|
| 1 | 2001 | 07 | 19 | +312 | 19.241 | 19.681 | 0.315 | 2.631 | 3.457 | 11.3 | 14.887 | 14.141 |
| 2 | 2001 | 08 | 10 | +334 | 19.293 | 19.733 | 0.529 | 2.567 | 3.555 | 4.3 | 14.932 | 14.648 |
| 3 | 2001 | 08 | 11 | +335 | 19.368 | 19.808 | 0.306 | 2.567 | 3.560 | 3.9 | 15.004 | 14.747 |
| 4 | 2001 | 08 | 12 | +336 | 19.290 | 19.730 | 0.469 | 2.568 | 3.564 | 3.6 | 14.922 | 14.684 |
| 5 | 2001 | 08 | 13 | +337 | 19.272 | 19.712 | 0.444 | 2.569 | 3.568 | 3.2 | 14.894 | 14.683 |
| 6 | 2001 | 09 | 21 | +376 | 20.419 | 20.859 | 0.791 | 2.866 | 3.718 | 9.3 | 15.721 | 15.107 |
| 7 | 2001 | 09 | 22 | +377 | 20.446 | 20.886 | 0.915 | 2.879 | 3.722 | 9.6 | 15.736 | 15.102 |
| 8 | 2001 | 09 | 23 | +378 | 20.149 | 20.589 | 0.331 | 2.893 | 3.725 | 9.8 | 15.427 | 14.780 |
| 9 | 2001 | 09 | 25 | +380 | 20.714 | 21.154 | 0.551 | 2.921 | 3.732 | 10.3 | 15.967 | 15.287 |
| 10 | 2001 | 10 | 06 | +391 | 21.241 | 21.681 | 0.534 | 3.091 | 3.769 | 12.4 | 16.349 | 15.531 |
| 11 | 2001 | 10 | 08 | +393 | 21.105 | 21.545 | 0.390 | 3.124 | 3.775 | 12.7 | 16.187 | 15.349 |
| 12 | 2002 | 09 | 10 | +730 | 19.559 | 19.999 | 0.293 | 2.953 | 3.959 | 0.8 | 14.660 | 14.607 |
| 13 | 2002 | 09 | 12 | +732 | 19.319 | 19.759 | 0.350 | 2.950 | 3.955 | 1.0 | 14.424 | 14.358 |
| 14 | 2002 | 09 | 13 | +733 | 19.341 | 19.781 | 0.320 | 2.949 | 3.953 | 1.2 | 14.448 | 14.369 |
| 15 | 2002 | 09 | 14 | +734 | 19.339 | 19.779 | 0.288 | 2.948 | 3.951 | 1.4 | 14.448 | 14.356 |
| 16 | 2002 | 09 | 15 | +735 | 19.365 | 19.805 | 0.351 | 2.948 | 3.949 | 1.7 | 14.475 | 14.363 |
| 17 | 2002 | 09 | 16 | +736 | 19.365 | 19.805 | 0.129 | 2.948 | 3.944 | 2.3 | 14.478 | 14.326 |
| 18 | 2002 | 10 | 03 | +753 | 19.78 | 20.19 | 0.34 | 3.020 | 3.925 | 7.1 | 14.81 | 14.34 |
| 19 | 2002 | 10 | 04 | +754 | 19.74 | 20.14 | 0.40 | 3.026 | 3.923 | 7.4 | 14.76 | 14.27 |

* YYYYMMDD= date; $\Delta t$= time after perihelion [d]; $<m(R)>$= mean observed magnitude;
$A_{PTV}$= Peak to valley amplitude; $m(V)$= visual magnitude = $m(R) - (V-R)$;
$\Delta$= comet-Earth distance [AU]; R= Sun-comet distance [AU]; $\alpha$ = phase angle [º];
$V(1,1,\alpha)$= absolute magnitude at $\alpha$ ; $V(1,1,0)$= absolute nuclear magnitude.

## 8. References

Barker, E.S., Cochran, A. L., Rybski, P.M., 1981. Observations of faint comets at McDonald Observatory: 1978-1980, pp. 150-155, In "Modern Observational Techniques for Comets", Brand, J.C., Greenberg, J.M., Donn, B., Rahe, J., Editors, JPL-Publication 81-68, 150-155.

Belton, M.J.S., Samarasinha, N. H., Fernandez, Y.R., Meech, K.J., 2005. The Excited Spin State of Comet 2P/Encke. Icarus, 175, 181-193.

Biesecker, D., 2000. Observations of comet 2P/Encke. ICQ, 22, p. 124-125, p. 142-143.

Cowan, J.J., A'Hearn, M.F., 1979. Vaporization of comet nuclei: Light curves and life times. Moon and Planets, 21, 155-171.

Delsemme, A.H., 1982. Chemical Composition of the Cometary Nucleus, 85-130, in "Comets", L. Wilkening, Editor, Univ. of Arizona Press, Tucson.

Fernandez, Y.R., Lisse, C.M., Kaufl, H.U., Peschke, S.B., Weaver, H.A., A'Hearn, M.F., Lamy, P.P., Livengood, T.A., Kostiuk, T., 2000. Physical Properties of the Nucleus of Comet 2P/Encke, Icarus, 147, 145-160.

Fernandez, Y.R., Lowry, S.C., Weissman, P.R., Mueller, B.A., Samarasinha, N.H., Belton, M.J.S., Meech, K.J., 2005. New Near-Aphelion Light Curves of Comet 2P/Encke. Icarus, 175, 194-214.

Ferrín, I., Gil, C., 1988. The aging of comets Halley and Encke. Astron. Astrophys., 194, 288-296.

Ferrín, I., 2005a. Secular Light Curve of Comet 28P/Neujmin 1, and of Comets Targets of Spacecraft, 1P/Halley, 9P/Tempel 1, 19P/Borrelly, 21P/Grigg-Skejellerup, 26P/Giacobinni-Zinner, 67P/Chruyumov-Gersimenko, 81P/Wild 2. Icarus 178, 493-516.

Ferrín, I., 2005b. Variable Aperture Correction Method in Cometary Photometry, ICQ 27, p. 249-255.

Ferrín, I., 2006. Secular Light Curve of Comets, II: 133P/Elst-Pizarro. Icarus, 185, 523-543.

Ferrín, I., 2007. Secular Light Curve of Comet 9P/Tempel 1. Icarus, 187, 326-331.

Ferrín, I., 2008. Atlas of Secular Light Curves of Comets. Submitted to Icarus.

Festou, M., C., Barale, O., 2000. The asymmetric coma of comets. I. Asymmetric outgassing from the nucleus of comet 2P/Encke. An. J., 119, 3119-3132.

Green, D.W., 2006. International Comet Quarterly.

Harmon, J. K., Campbell, D. B., Ostro, S.J., Nolan, M.C., (1999). Radar observations of comets, Planetary and Space Science, 47, p. 1409-1422.

Harmon, J.K., Nolan, M.C., 2005. Radar observations of Comet 2P/Encke during the 2003 apparition. Icaus, 176, 175-183.

Hergenrother, C.W., Mueller, B.A., Campins, H., Samarasinha, N.H., McCarthy, D.W., 2006, R and J-band photometry of Comets 2P/Encke and 9P/Tempel 1. Icarus, 181, 156-161.

Ho Peng Yoke, 1962. Ancient and Mediaeval observations of comets and novae in Chinese sources. Vistas in Astronomy, 5, 127-225.

Jewitt, D., Meech, K., 1987. CCD Photometry of Comet P/Encke. An. J., 93, 1542-1548.

Kamel, L., 1991. The evolution of P/Encke's light curve: No secular fading, a vanishing perihelion asymmetry. Icarus, 93, 226-245.

Kamel, L., 1992. The comet light curve atlas. Astron. and Astrophys., 92, 85-149.

Kamoun, P.G., Campbell, D.B., Ostro, S.J., Pettengil, G.H., Shapiro, I.I., 1981. Comet Encke: Radar detection of nucleus. Science, 216, 293-295.

Kelley, M.S., Woodward, C.E., Harker, D.E., Wooden, D.H., Gehrtz, R.D., Campins, H.,

Hanner, M.S., Lederer, S.M., Osip, D.S., Pittichova, J., Polomski, E., 2006. arXiv:astro-ph/0607416v1, 18 Jul 2006, 1-48.

Levison, H.F., Duncan, M.J., 1997.  From the Kuiper Belt to the Jupiter family of comets: The spatial distribution of ecliptic comets.  Icarus, 127, 13-32.

Levison, H.F., Terrell,D., Wiegert, P.A., Dones, L., Duncan, M.J., 2006.   On the origin of the unusual orbit of comet 2P/Encke.  Icarus, 182, 161-168.

Lamy, P., Biesecker, D.A., Groussin, O., 2003.  SOHO /LASCO observations of an outburst of Comet 2P/Encke at its 2000  perihelion passage.  Icarus, 163, 142-149.

Lamy, P.L., Toth, I., Fernandez, Y.R., Weaver, H.A., 2005.  The Sizes, Shapes, Albedos and Colors of Cometary Nuclei.  223-264, In Comets II, Festou, M., Keller, H.U., Weaver, H. E., Editors, Univ. of Arizona Press, Tucson.

Lowry, S.C., Weissman, P.R., Sykes, M.V., Read, W.T., 2003.  Observations of Periodic Comet 2P/Encke: Physical Properties of the Nucleus and First Visual-Wavelength Detection of its Dust Trail.  Lunar and Planetary Science, XXXIV, 2956-2057.

Lowry, S.C., Weissman, P. R., 2007.  Rotation and color properties of the nucleus of Comet 2P/Encke.  Icarus, 188, p. 212-223.

Luu, J. Jewiit, D., 1990.  The nucleus of comet 2P/Encke.  Icarus, 86, 69-81.

Lüthen, H., Ferrín, I., Green, D.W.E., Bortle, J.E., 2000.  Max Beyer (1894-1982): A Master of Comet Observing.  ICQ, 22, 105-114.

Meech, K.J., Svoren, J., 2005.  Using Cometary Activity to Trace the Physical and Chemical Evolution of Cometary Nuclei.  317-335, In Comets II, Festou, M., Keller, H.U., Weaver, H. E., Editors, Univ. of Arizona Press, Tucson.

Meech, K.J., Fernandez, Y., Pittichova, J., 2001.  Aphelion activity of 2P/Encke.  Bull. Amer. Astron. Soc., 33, #20.06

Meech, K.J., Hainaut, O.R., Marsden, B.G., 2004.  Comet nucleus size distributions from HST and Keck telescopes.  Icarus, 170, 463-491.

Ney, E.P., 1974.  Multiband photometry of comets Kohoutek, Bennett, Bradfield and Encke.  Icarus, 23, 551-560.

Pittich, E.M., D'Abramo, G., Valsecchi, G.B., 2004.  From Jupiter-family to Encke-like orbits.  Astron. and Astrophys., 422, 369-375.

Reach, W.T., Sykes, M.V., Lien, D., Davies, J.K., 2000.  The formation of Encke Meteoroids and Dust Trail.  Icarus, 148, 80-94.

Scholl, H., Froeschle, Ch., 1991.  The   secular resonance region near 2 AU: a possible source of meteorites.  Astron. Astrophys., 245, 316-321.

Samarasinha, N.H., 1997.  Preferred orientation for the rotational angular momentum vector of periodic comets.  American Astronomical Society, DPS Meeting #28, BAAS, 29, 743.

Sekanina, Z., 1976. A continuing controversy: Has the cometary nucleus been resolved ?  In: The Study of Comets, Donn, B., Mumma, M., Jackson, W., A'Hearn, M., Harrington, R., Editors, NASA, Washington, 537-585.

Sekanina, Z., 1988a.  Outgassing asymmetry of periodic comet Encke I.  Apparitions 1924-1984.  An.J., 95, 911-924.

Sekanina, Z., 1988b.  Outgassing asymmetry of periodic comet Encke II.  Apparitions 1868-1918.  An.J., 96, 1455-1475.

Weissman, P. R., 1980.  Physical Loss of Long-period comets.  Astron. Astrophys. 85, 191-196.

Whipple, F., Hamid, S.E., A search for Encke´s comet  in ancient Chinese records:

A progress report.  In G. A. Chebotarev, E. I. Kazimirchak-Polonskaya, and B. G. Marsden (Eds), The Motion, Evolution of Orbits, and Origin of Comets, Reidel, Dordrecht, 152-154.

Whipple, F., Sekanina, Z., 1979.  Comet Encke:  Precession of the spin axis, non-gravitational motion, and sublimation.  An. J., 84, 1894-1909.

Zissell, R. E., Transformation of AAVSO Archive Visual Data to Johnson V System. JAAVSO, 31, 128-137.

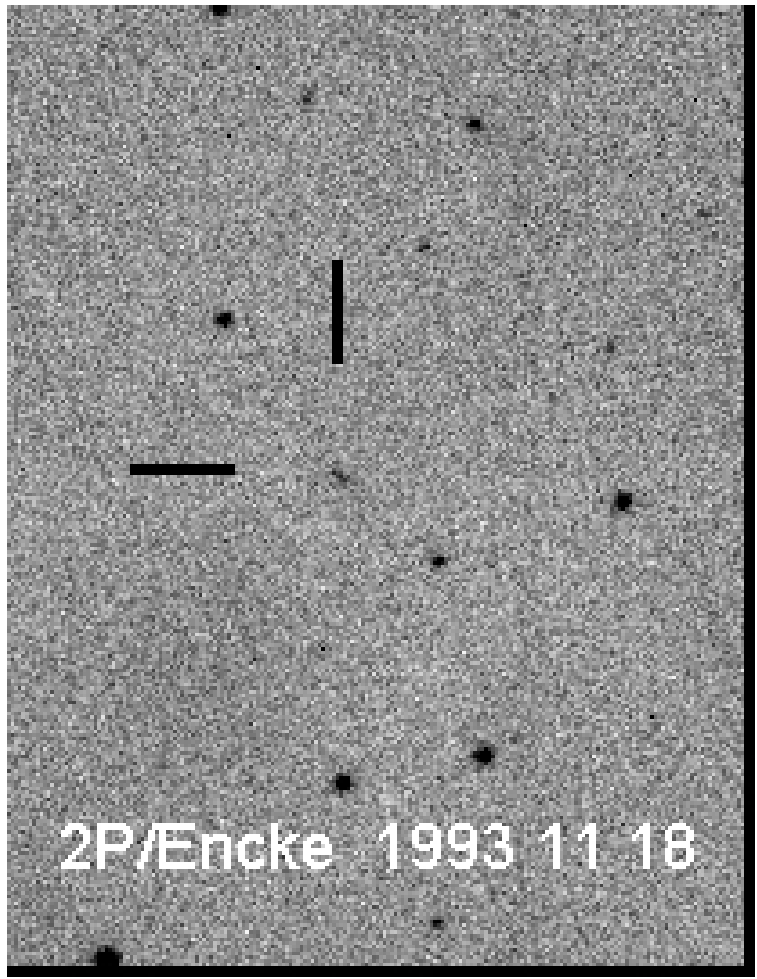

2P/Encke 1993 11 18

Figure 1

Figure 1.  Trailed image of comet 2P/Encke, taken with the 1 m f/3 Schmidt telescope of the National Observatory of Venezuela.  The date was November 18[th] , 1993 and the CCD was unfiltered.   Exposure time, 420 s.  The comet was at $\Delta$= 0.94 AU from the Earth, and R= 1.60 AU from the Sun, phase angle $\alpha$= 35.6°.  The image was taken 83 days before perihelion and +4 days after turn on of the comet.   Perihelion time, $T_{PERI}$ = 1994, February 9[th].   The magnitudes were  R($\Delta$,R,$\alpha$)=17.0±0.15,   V($\Delta$,R,$\alpha$)=17.44±0.15,   R(1,1,$\alpha$)= 16.11±0.15, R(1,1,0)= 15.08±0.15.  The image is 5'x6.7' in size.  There is an unresolved coma not detected in this rendition.

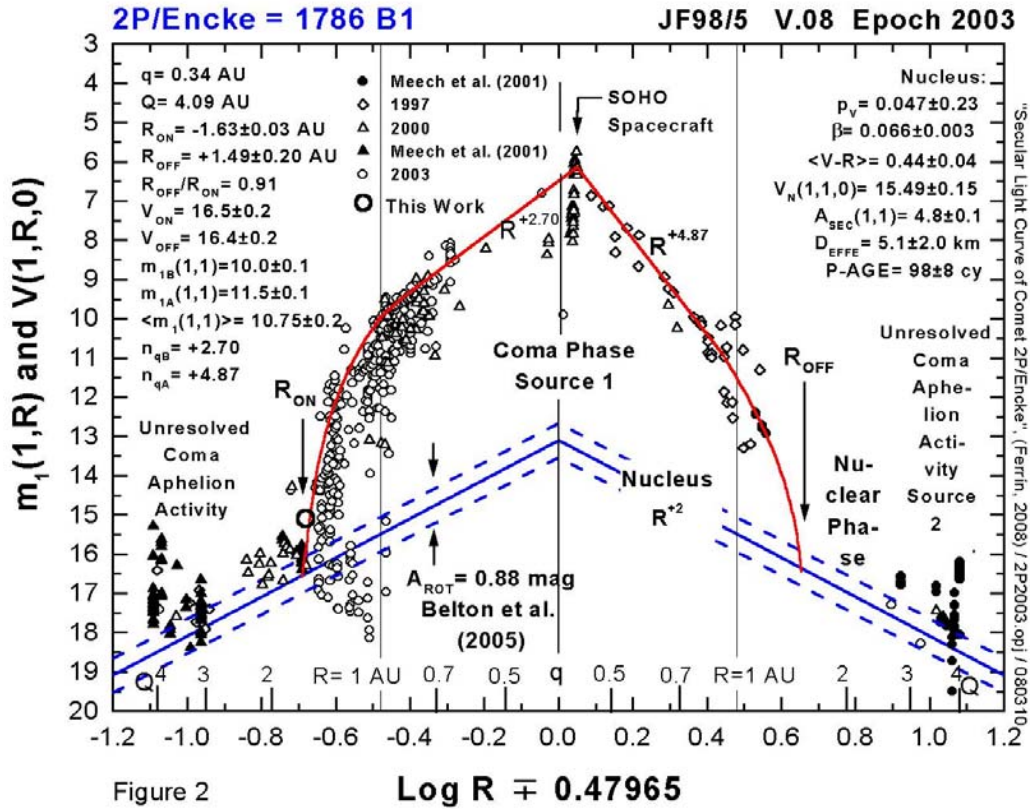

Figure 2

Figure 2. Secular light curve of comet 2P/Encke, log plot. The x-axis has been increased in Log q = ±0.47965 to make room for observations inside the Earth's orbit. The slope 5 line at the bottom of the plot in the form of a pyramid is due to the atmosphereless nucleus. Notice the sharp turn on point, and the activity at aphelion evidenced by observations well above the nuclear line. Although the SLC exhibits almost the same turn on and turn off distances, the SLC is asymmetric at 1 AU from the Sun, as evidenced by the fact that the absolute magnitude before perihelion, $m_{1B}(1,1)$= 9.8±0.1, while after perihelion is $m_{1A}$ (1,1)= 11.5±0.1, an asymmetry of 1.7±0.15 magnitudes at 1 AU. The photometric age derived from this plot is P-AGE= 97±8 cy (comet years), corresponding to an old comet. The mean value <V-R> and $p_R$ are taken from the literature cited in the text. Our observation was taken +4 days after turn on.

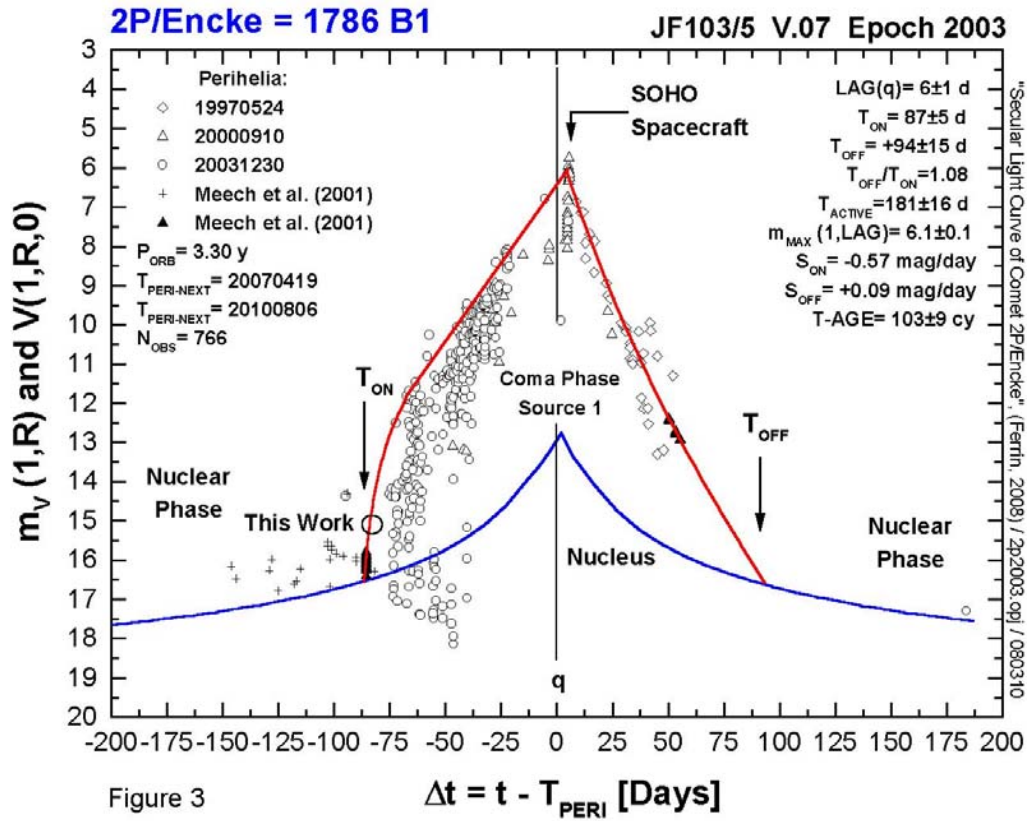

Figure 3

Figure 3. Secular light curve of comet 2P/Encke, time plot, Source 1.   Notice the time delay between maximum brightness and perihelion, LAG.   The time-age derived from this plot is T-AGE= 103±9 cy, corresponding to an methuselah comet (defined as any comet with 100 cy < P-AGE).

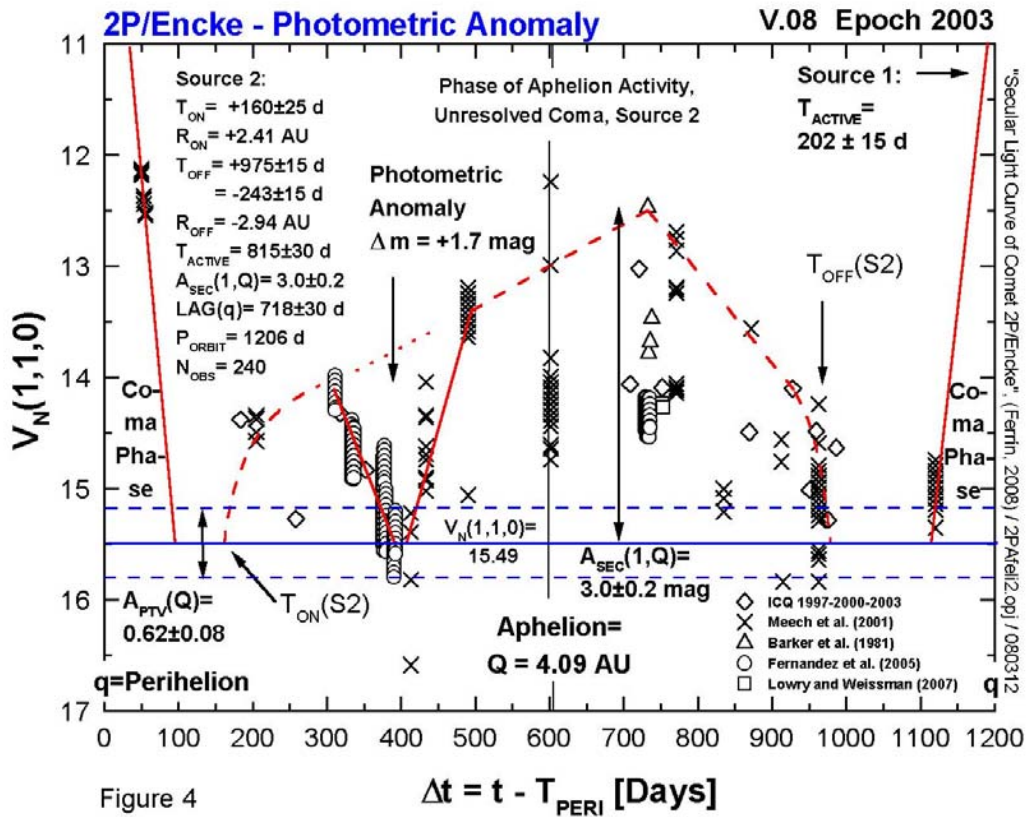

Figure 4. Secular light curve of comet 2P/Encke, time plot, activity at aphelion, Source 2. The dashed line tries to describe an envelope of the observations but in no way implies that the activity is sustained. Between $\Delta t$ = +310 and +392 d there is a well observed linear decay in brightness that might be indication of topography and is enlarged in Figure 5. The maximum activity (3.0±0.2 mag) is quite significant. The nuclear magnitude derived from the phase plot (Figure 6), is in agreement with the lowest observations at aphelion by Fernandez et al. (2005) and Meech et al. (2001) at $\Delta t$= +390 days.

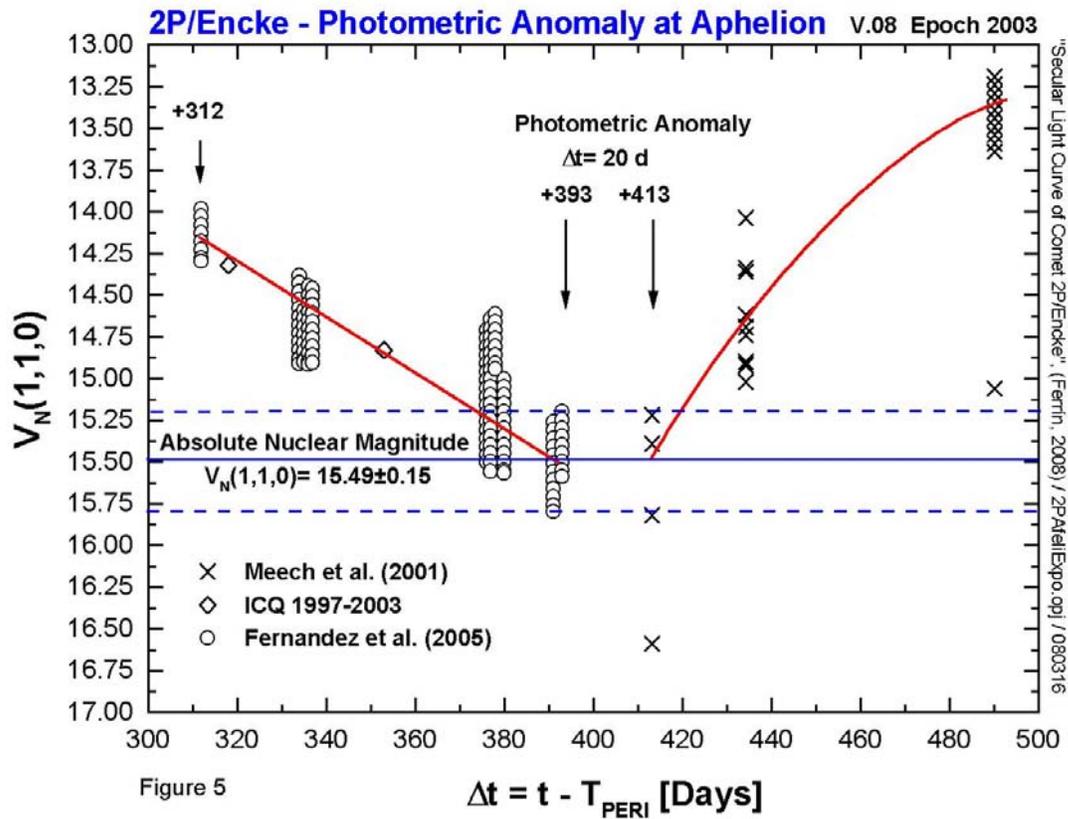

Figure 5. Comet 2P/Encke, photometric anomaly at aphelion. The anomaly has been expanded from Figure 4 to see its extent and shape. The brightness decreases linearly from $\Delta t$ = +310 to +392 d after perihelion, an interval of 82 days. There seems to be an absolute minimum of the secular light curve at $\Delta t$= +393 to 413. The photometric anomaly can be due to a topographic effect or the turn off of an active region. Additional observations of this region are highly desirable at every apparition because this might be the absolute minimum brightness of the whole orbit (not considering the rotational component).

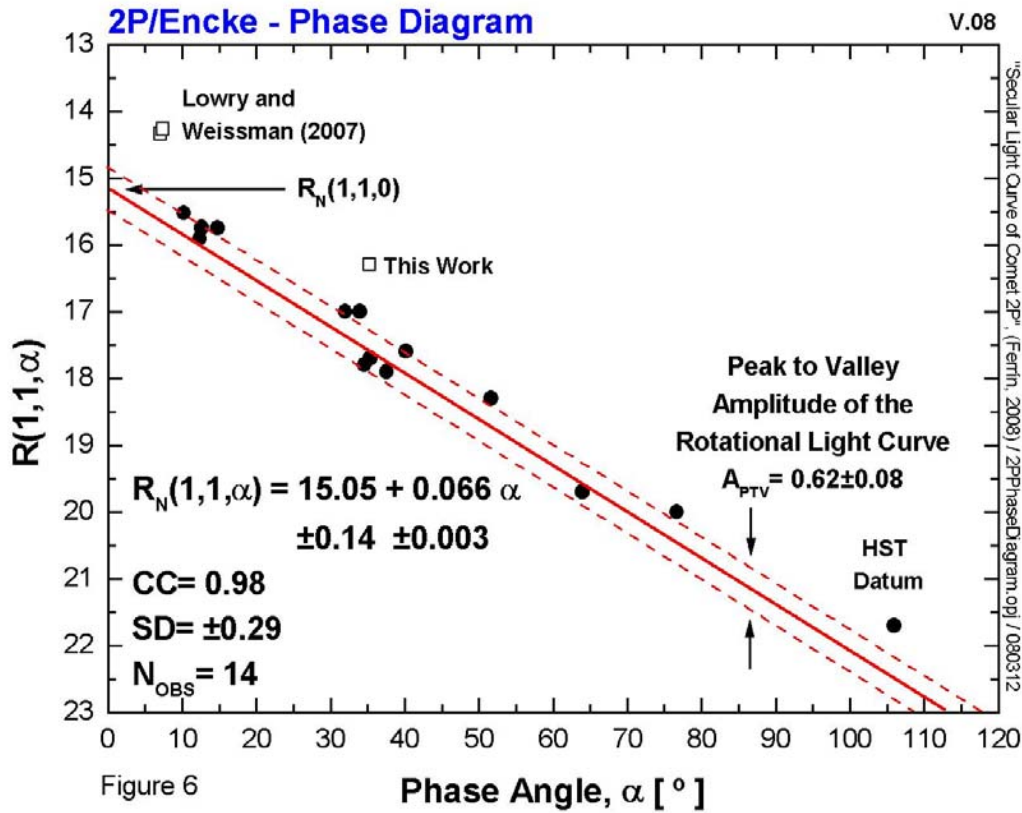

Figure 6. Phase diagram. The best way to determine the absolute magnitude of a comet is using the phase diagram. The magnitude at phase angle $\alpha= 0°$, is the absolute magnitude $R_N(1,1,0)$. To diminish the errors and to increase the robustness of this determination, we have imposed three new constraints: a) The observations must be devoid of coma (this implies that photographic observations are not included because of CN contamination line at 3883 A). b) The observations should lie inside the rotational amplitude of the nucleus. c) The absolute nuclear magnitude $R_N(1,1,0)$ must be in agreement with the faintest observations made at aphelion. In this way we determine $R_N(1,1,0)= 15.05\pm0.14$, $\beta= 0.066\pm0.003$. These values are in excellent agreement with previous values, but exhibit smaller errors. The standard deviation (SD) of the observations is in excellent agreement with the amplitude of the *rotational* light curve. The correlation coefficient (CC) is very large. This plot is also excellent to determine coma contamination (values well above the phase line). We see that our observation was already coma contaminated in spite of having been made only +4 days after turn on. However the coma is not apparent in the image (Figure 1). The photometry of Lowry and Weissman (2007) is also compromised by an unresolved coma. Notice the extent of the phase angle observations that reach to 105°, the largest of any comet

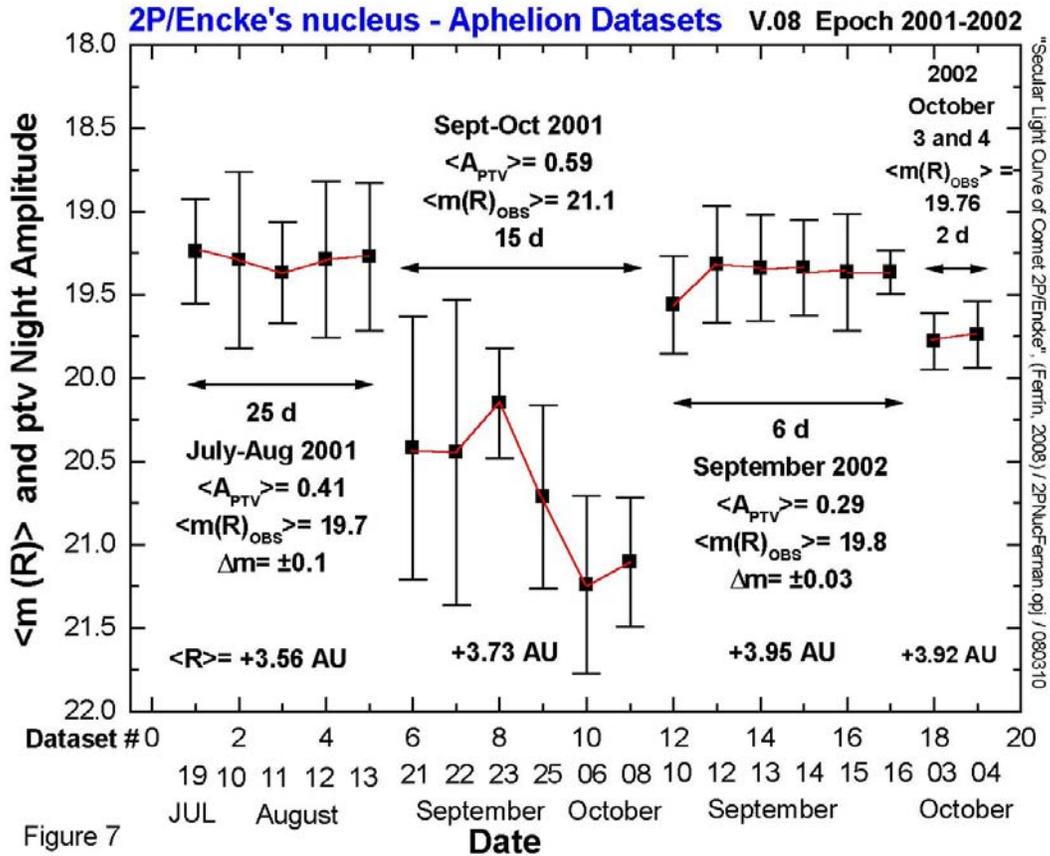

Figure 7. Data sets of Fernandez et al. (2005) and of Lowry and Weissman (2007) of *rotational* measurements. Means and ranges (vertical bars) have been calculated for every night, and plotted as a function of time. PTV= peak to valley. These observations have been made near aphelion, as can be deduced from the scale below, showing solar distances of +3.56 to +3.95 AU. There are four observing intervals. In the first, third and fourth interval (duration 25, 6 and 2 d) the comet remains practically constant in brightness within ±0.1, ±0.03 and ±0.04 magnitudes. The second interval in the middle (lasting 15 d) is much fainter than the other two and exhibits a drop of 1.1 magnitudes in 13 days. The lowest night is 1.5 magnitudes below the steady state nights. Notice also that the amplitude of the rotational light curve is larger for the fainter magnitudes, implying that brighter magnitudes are contaminated by an unresolved coma

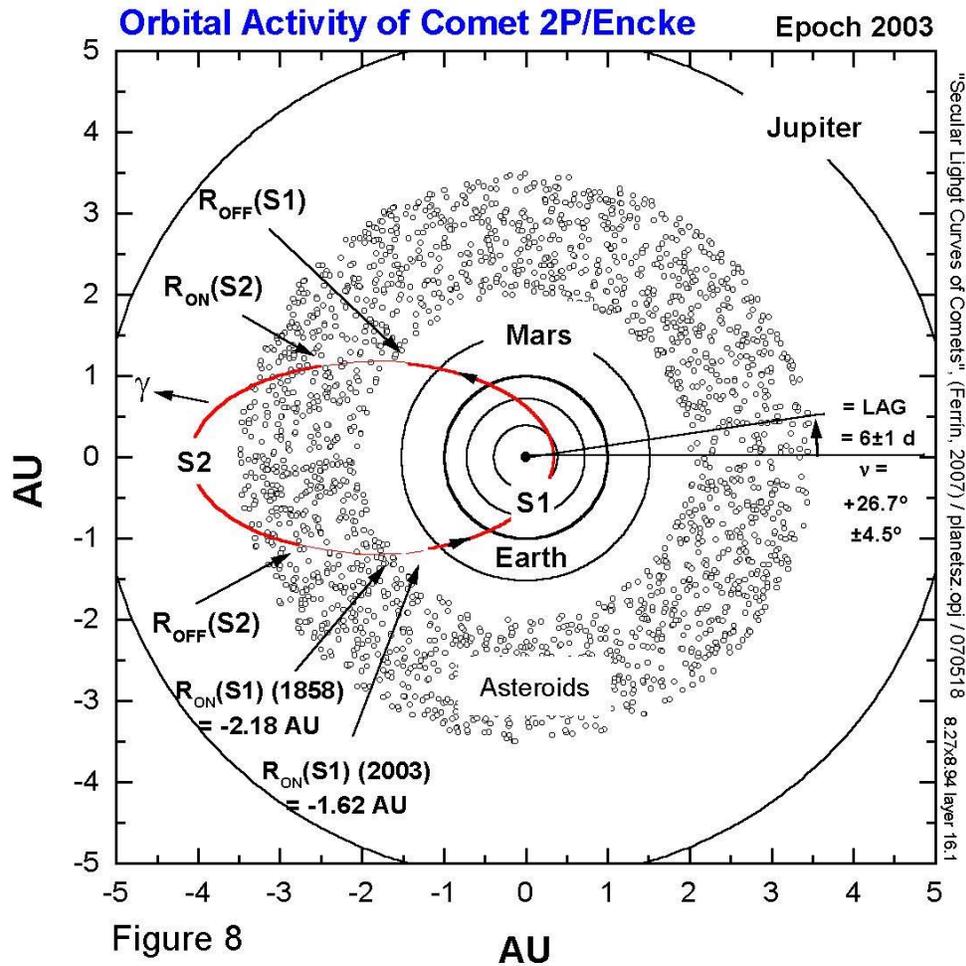

Figure 8. The orbit and activity of comet 2P/Encke, seen from above the orbital plane. The heavy line shows the regions of activity, Source 1 at perihelion, Source 2 at aphelion. Notice the turn on of activity in 1858, and 2003. The maximum activity takes place at LAG= 6±1 d after perihelion, which can be translated into a true anomaly of +26.7±4.5°.

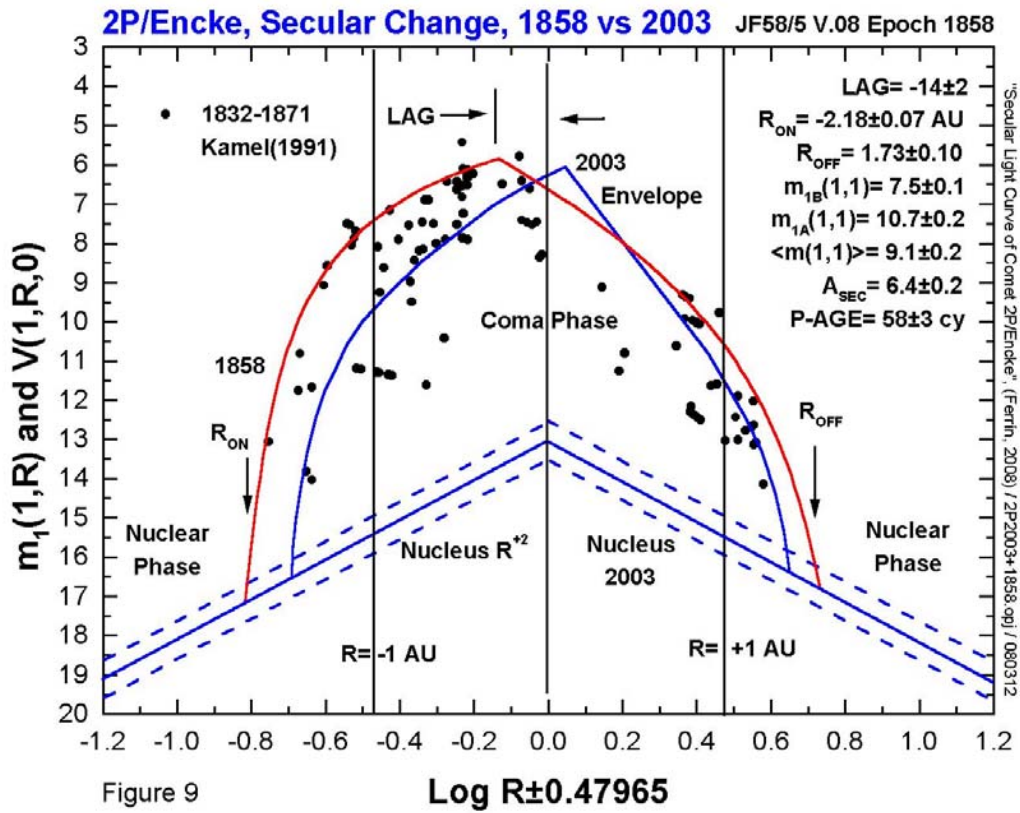

Figure 9. The SLC from 1858 compared with the SLC of 2003, Log plot. The data has been taken from Kamel (1992). The points are the observed brightness. The SLC has been narrowing with time, a result that is in accord with the time evolution of the comets presented in papers I to III. It is possible to derive a photometric age for this Epoch. Compare with Figure 2. Seven physical parameters show evolution with respect to 2003

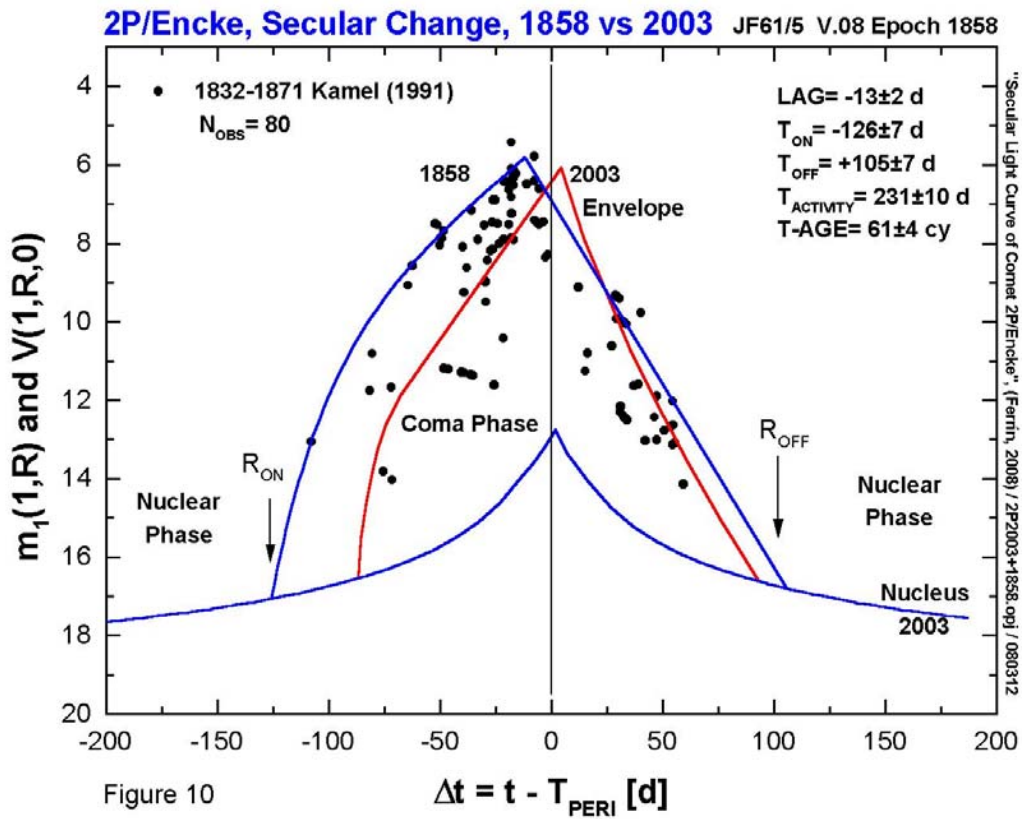

Figure 10. SLC from 1858 compared with the SLC of 2003, time plot. It can be ascertained that the SLC changed shape. The SLC has been narrowing with time, a result that is in accord with the time evolution of the comets presented in papers I to III. Compare with Figure 3.

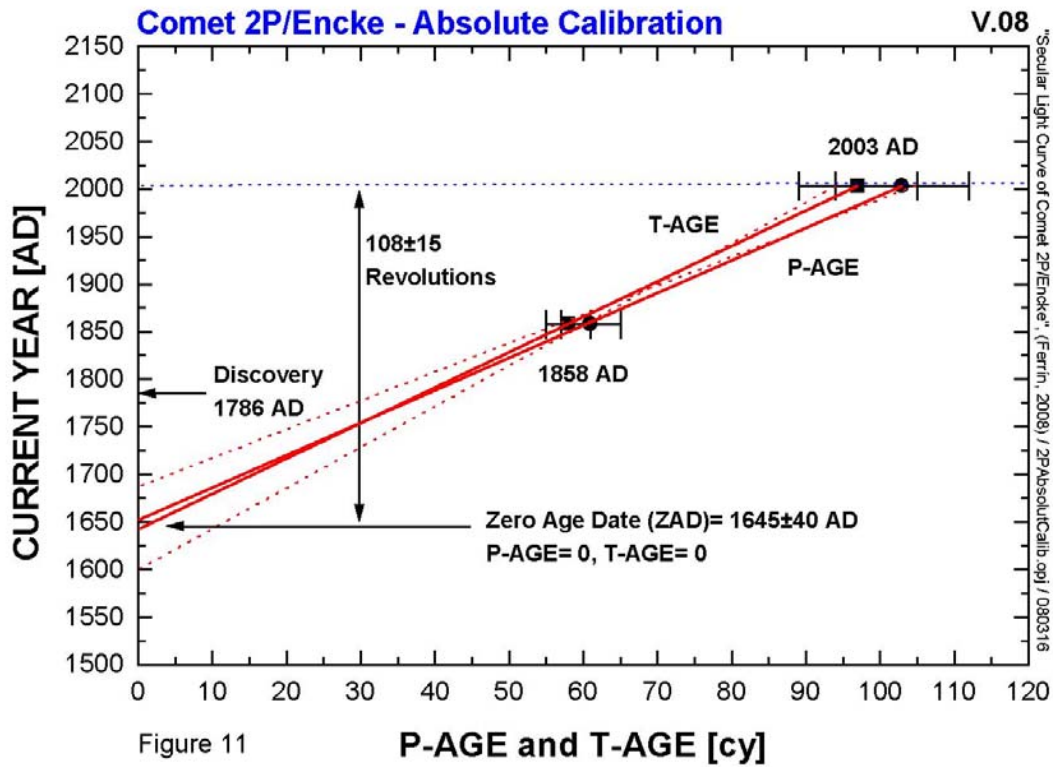

Figure 11.  Absolute calibration of age.   It can be seen that the two ages, P-AGE and T-AGE, give practically the same zero-age-date, ZAD = 1645±40 AD.    This is not the birth date of the comet, but the effective time at which it began sublimating.   However the final meaning of this ZAD is still a matter of interpretation

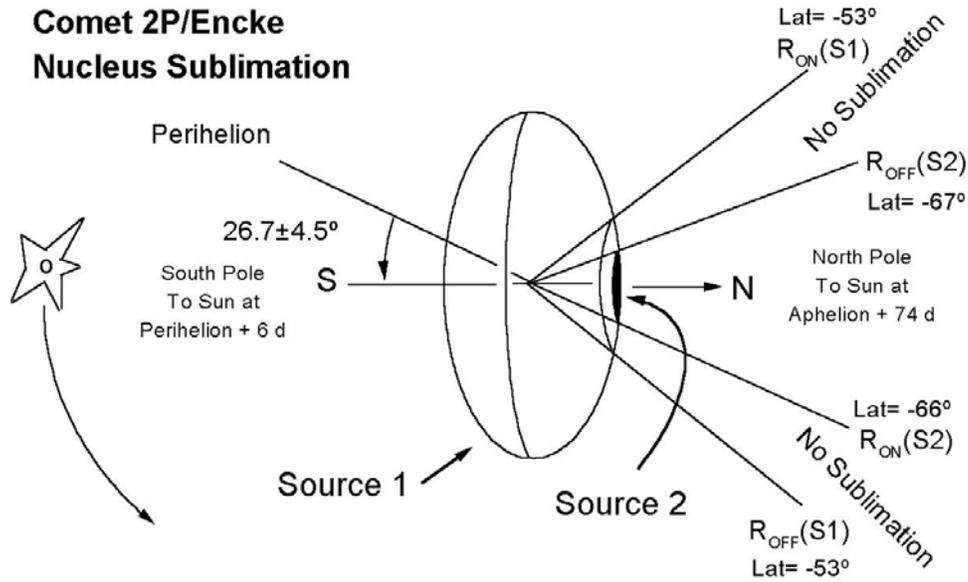

Figure 12.   The distribution of volatile ices over the nucleus of comet 2P/Encke can be deduced from Figure 8, if the pole is near the orbital plane.  It can be seen that source 1 occupies the southern hemisphere, while source 2 occupies the north pole.  However due to the fact that Source 2 is active at aphelion, the active time is much larger than for Source 1.    In the context of this investigation "a source" may include several active regions lumped together rather than a surface sublimating on every square inch.

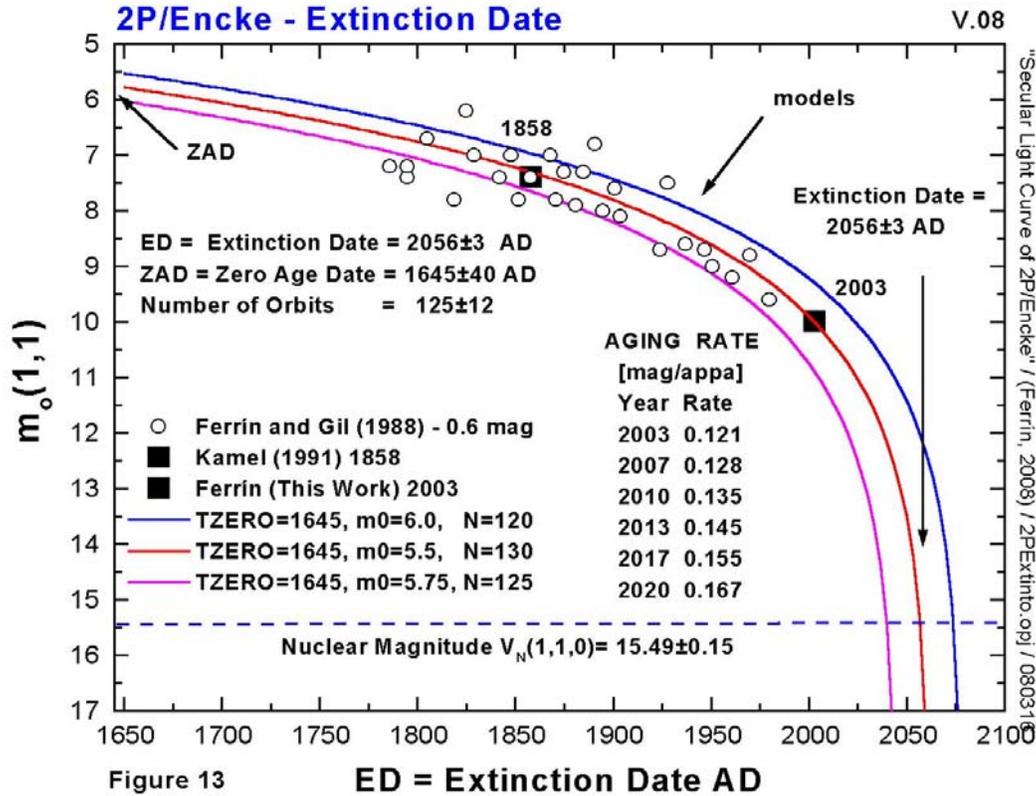

Figure 13. Extinction date of comet 2P/Encke. The secular absolute magnitudes have been plotted as a function of time and fitted with a suitable theoretical model assuming that the comet dies by sublimation. The original data set has been moved upward by -0.6 mag because the authors Ferrin and Gil (1988) took the mean values of the observations, while we now know that the envelope is the correct interpretation. Notice that with this correction, the absolute magnitude of the comet in the data sets of Ferrín and Gil and Kamel, coincide in 1858. An extinction date can then be deduced, and we find that ED = 2056±3 AD. This implies a total number of 125±12 revolutions since it entered the inner solar system. Notice the narrow range of models that fit the observations. The aging rate per apparition is accelerating and reaching a value easily measured by photometry. Thus this is a useful model because it can be tested.